\documentclass{ieeeaccess}
\usepackage{cite}
\usepackage{amsmath,amssymb,amsfonts}
\usepackage{algorithmic}
\usepackage{graphicx}
\usepackage{subfigure}
\usepackage{textcomp}
\usepackage{xcolor}
\usepackage{threeparttable}
\usepackage{footnote}
\usepackage{multirow}
\usepackage{url}

\def\BibTeX{{\rm B\kern-.05em{\sc i\kern-.025em b}\kern-.08em
    T\kern-.1667em\lower.7ex\hbox{E}\kern-.125emX}}

\newcommand{\tabincell}[2]{\begin{tabular}{@{}#1@{}}#2\end{tabular}}

\graphicspath{{fig/}}
\usepackage[caption=false,font=footnotesize]{subfig}

\begin{document}
\bstctlcite{IEEEexample:BSTcontrol}

\history{Received June 17, 2021, accepted June 29, 2021.}
\doi{10.1109/ACCESS.2021.3096664}

\title{A Deep Learning-Based FPGA Function Block Detection Method with Bitstream to Image Transformation}
\author{\uppercase{Minzhen~Chen~and~Peng~Liu},~\IEEEmembership{Member, IEEE}}
\address{College of Information Science and Electronic Engineering, Zhejiang University, Hangzhou 310027, China}

\markboth
{Chen \headeretal: A Deep Learning-Based FPGA Function Block Detection Method with Bitstream to Image Transformation}
{Chen \headeretal: A Deep Learning-Based FPGA Function Block Detection Method with Bitstream to Image Transformation}

\corresp{Corresponding author: Peng Liu (e-mail: liupeng@zju.edu.cn).}

\begin{abstract}
In the context of various application scenarios and/or for the sake of strengthening field-programmable gate array (FPGA) security, the system functions of an FPGA design need to be analyzed, which can be achieved by systematically partitioning the FPGA's bitstream into manageable functional blocks and detecting their functionalities thereafter.
In this paper, we propose a novel deep learning-based FPGA function block detection method with three major steps.
In specific, we first analyze the format of the bitstream to obtain the mapping relationship between the configuration bits and configurable logic blocks because of the discontinuity of the configuration bits in the bitstream for one element.
In order to reap the maturity of object detection techniques based on deep learning, our next step is to convert an FPGA bitstream to an image, following the proposed transformation method that takes account of both the adjacency nature of the programmable logic and the high degree of redundancy of configuration information.
Once the image is obtained, a deep learning-based object detection algorithm is applied to this transformed image, and the objects detected can be reflected back to determine the function blocks of the original FPGA design. The deep neural network used for function block detection is trained and validated with a specially crafted bitstream/image dataset.
Experiments have confirmed high detection accuracy of the proposed function detection method, showing a 98.11\% of mean Average Precision (IoU=0.5) for 10 function blocks within a YOLOv3 detector implemented on Xilinx Zynq-7000 SoCs and Zynq UltraScale+ MPSoCs.

\end{abstract}

\begin{keywords}
Bitstream-to-image transformation, field-programmable gate array, function block detection.
\end{keywords}

\titlepgskip=-15pt

\maketitle

\section{Introduction}\label{1_sec_introduction}

\PARstart{F}{ield-programmable} gate arrays (FPGAs) are gaining prominence in a wide array of fields, such as communication, deep learning, and digital signal processing, due to their distinct features and advantages of configurability, fast development cycle, and availability of abundant logic/storage resources. Along with the widespread adoption of FPGA technologies comes a pressing security concern of the FPGA-based systems. Since the functions and behaviors of an FPGA design are fully determined by its bitstream that contains the configuration data to be loaded into the FPGA during power-on, the bitstream is becoming a weak part of the FPGA security. One way to fully assess the resilience of an FPGA design boils down reverse-engineering the bitstream to detect/isolate various function blocks and correspondingly determine their intended functionalities.

\begin{figure}[!t]
\centerline{\includegraphics[width=3.5in]{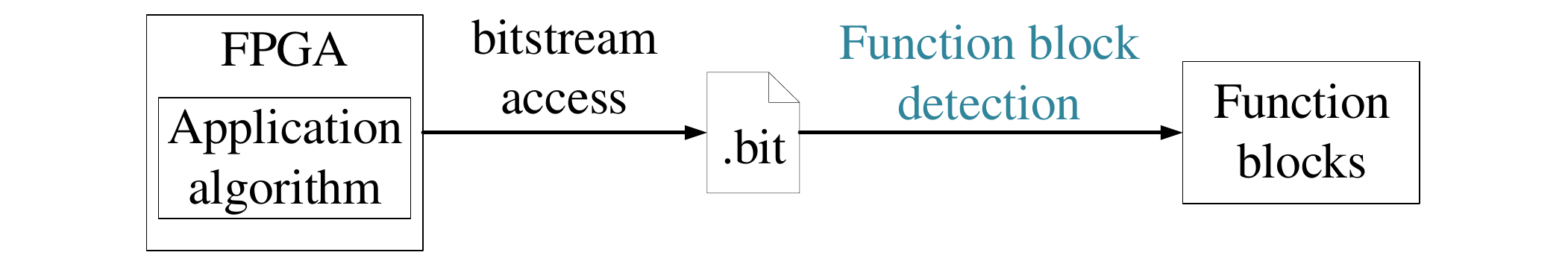}}
\caption{For instance, an application scenario demonstrates that FPGA function block detection helps analyze the circuit's system function.}
\label{fig1}
\end{figure}

A function block in an FPGA system design refers to a circuit block implementing a complex function, such as a cryptographic operator like MD5 (Message Digest Algorithm 5)~\cite{MD5_1992}. The application algorithm implemented on an FPGA contains different kinds of function blocks. For example, a typical FPGA implementation of the encryption algorithm used for PDF-R2 contains two kinds of function blocks, namely MD5 and RC4 (Rivest Cipher 4)~\cite{RC4_1994}. Function block detection plays a role of pre-analysis in the analysis of the system function, as shown in Fig.~\ref{fig1}.

One approach to defining the function blocks of an FPGA design is through circuit partitioning with the circuit represented directly by its bitstreams or netlists, after which the content of the partitioned circuits will be compared against the existing designs~\cite{Ziener_FPL2006,Couch_HOST2016}. One drawback of such an approach is also attributed to the time-consuming partitioning process that often fails to create a perfect partition of the circuit, which can lead to incorrect matching results. Features used to compare the partitioned circuits need to be designed manually, and improperly selected features result in the performance degradation of the conventional approaches.

In this paper, we propose an FPGA function block detection method that is able to detect one or more function blocks in the FPGA design from a given complete bitstream.
The bitstream-to-image transformation turns the FPGA function block detection problem into an image objection detection task, which has been solved well by deep learning methods. The image transformed from the bitstream should reflect the adjacency between the elements in the programmable logic.
There are two major issues that complicate the critical bitstream-to-image transformation: 1) a high degree of redundancy of configuration information and 2) discontinuity of configuration bits for one element in the bitstream. To address these issues, only the configuration bits of the configurable logic blocks (CLBs) are used for transformation and the mapping relationship between the configuration bits and CLBs is found.

In summary, the main contributions of this paper include:
\begin{itemize}
\item The bitstream-to-image transformation suitable for deep learning processing is proposed by analyzing the mapping relationship between the configuration bits and CLB elements.
\item A dataset, in which the images are transformed from bitstream files containing 10 kinds of cryptographic operators, is generated for deep learning without manual annotation, which means there is no need to label the data by humans.
\item The deep learning techniques are applied to FPGA function block detection from bitstream for the first time by training a deep learning-based object detection algorithm on the dataset.
    A number of FPGA designs for application-specific encryption algorithms~\cite{Liu_TC2019} have been adopted to test and validate the proposed detection method. It is found that the function blocks of the FPGA designs can be successfully identified.
    The mean Average Precision (mAP) reaches 98.11\% for 10 kinds of function blocks when Intersection over Union (IoU) is 0.5.
\end{itemize}

The rest of the paper is organized as follows.
Section~\ref{5_sec_related_works} discusses the related works. Section~\ref{3_sec_Methodology} firstly introduces the overall process of the detection method and then describes each step of the method in detail. The experimental results are presented in Section~\ref{4_sec_Experimental_results} and Section~\ref{6_sec_conclusions} summarizes the findings.

\section{Related Works}\label{5_sec_related_works}

In this section, some related works about bitstream format analysis, bitstream reverse engineering, and deep learning-based circuit classification are presented.

An FPGA bitstream contains the programming information configuring the programmable logic in the FPGA. Due to the lack of disclosed information about the bitstream format from FPGA vendors, many works have analyzed the format of bitstream~\cite{Ziener_FPL2006,leRoux2019,DangPham_DATE2017_BITMAN,Bozzoli_ARCSW2018_COMET}. Ziener \emph{et al.}~\cite{Ziener_FPL2006} extracted the content of Look Up Tables (LUTs) in the bitstream of Xilinx Virtex-II and Virtex-II Pro FPGAs to identify intellectual property (IP) cores in the FPGAs. Le Roux \emph{et al.}~\cite{leRoux2019} analyzed the bitstream of Xilinx Virtex-5 FPGAs to manipulate the configuration bits of LUTs for the purpose of reconfiguring the FPGAs in real-time. There are also some related works analyzing the bitstream format of the later Xilinx 7-series FPGAs~\cite{DangPham_DATE2017_BITMAN,Bozzoli_ARCSW2018_COMET}. Dang Pham \emph{et al.}~\cite{DangPham_DATE2017_BITMAN} provided a tool called BITMAN that supports bitstream manipulations, such as module placement, module relocation, and so on. COMET~\cite{Bozzoli_ARCSW2018_COMET} is a tool supporting bitstream analysis, visualization, and manipulation. The manipulation of bitstream is to provide means to perform partial reconfiguration or fault injection. The bitstream brings up lots of security issues for FPGAs, such as bitstream access, bitstream decryption~\cite{Moradi_CCS2011,Tajik_CCS2017,Ender_USENIXsecurity2020}, and bitstream reverse engineering~\cite{note_FPGA2008_bitstream,Benz_FPL2012,DING_2013_NCD,Zhang_Access_2019}.

There are many works implementing FPGA bitstream reverse engineering based on bitstream analysis. Some of them reverse the bitstream to a Xilinx Design Language (XDL) level representation of the netlist or a Native Circuit Description (NCD) file~\cite{note_FPGA2008_bitstream,Benz_FPL2012,DING_2013_NCD}. Some of them further reverse the netlist file to Register Transfer Level (RTL) code~\cite{Zhang_Access_2019}. These works analyze the bitstream format and gather databases containing the mapping relationship from the configuration bits in the bitstream to their related configurable elements.
These databases are used to reverse engineer. Bitstream reverse engineering aims at performing analysis or detecting hardware Trojan based on netlist or RTL code.

Dai \emph{et al.}~\cite{Dai_HOST2017} applied a convolutional neural network (CNN) method for arithmetic operator classification and detection from gate-level circuits, and discussed the importance of representation of circuits for CNN processing.
Fayyazi \emph{et al.}~\cite{Fayyazi_DATE2019} also presented a CNN-based gate-level circuit recognition method and used a vector-based representation for the CNN processing. The method can be used to detect the hardware trojans or classify the arithmetic operators, such as an adder and a multiplier.
Gate-level circuit recognition requires the bitstream
to be reverse engineered first.
Mahmood \emph{et al.}~\cite{Mahmood_ICCE_Berlin2019} proposed judging whether a partial bitstream contains a hardware module implementing an add operation using neural networks.
Neto \emph{et al.}~\cite{Neto_ICCAD2019} took advantage of the latest progress of deep neural networks (DNNs) in image classification to choose the best optimization method for the partitioned circuits. A binary image-like representation is proposed to represent Boolean logic function, in which the white pixel represents the logic true (1), and the black pixel represents the logic false (0). However, the binary image in our work is used to represent the utilization of the slices.

\section{Proposed Function Detection Method through Bitstream to Image Transformation}
\label{3_sec_Methodology}

\begin{figure*}[!t]
\centerline{\includegraphics[width=7.1in]{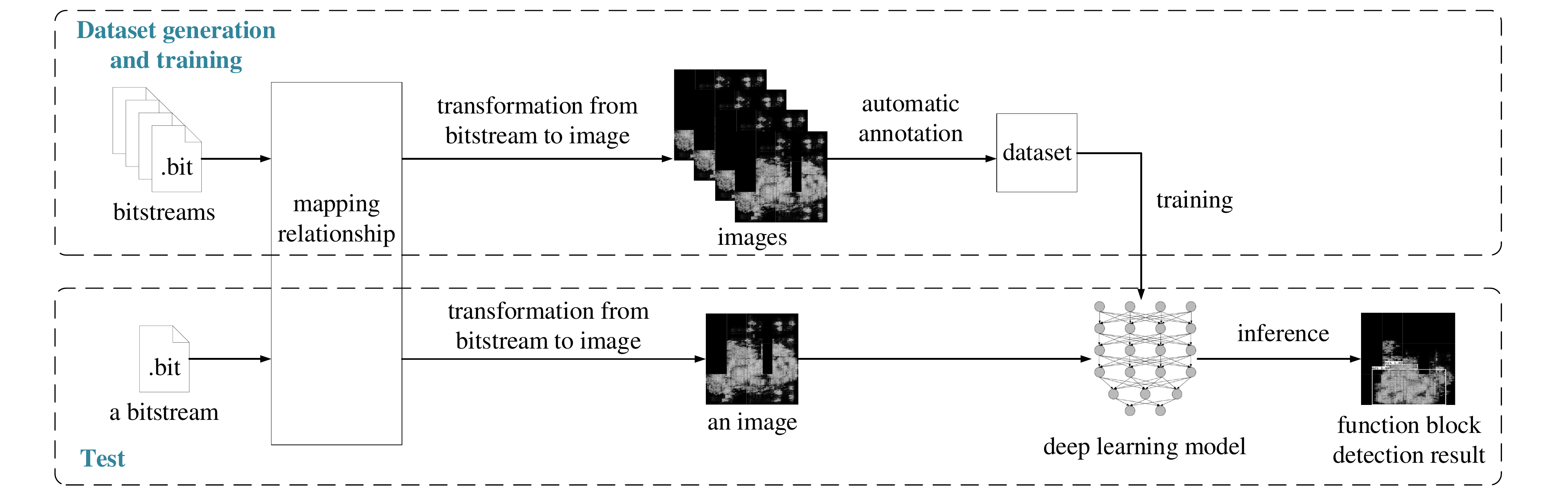}}
\caption{The process of function block detection method consisting of the transformation from bitstream to image, dataset generation, deep learning training, and deep learning inference.}
\label{fig2}
\end{figure*}

The process of function block detection consists of the following steps, as is shown in Fig.~\ref{fig2}. First of all, the transformation from bitstream to image is proposed by analyzing the mapping relationship between the configuration bits and CLBs. Then, the image will be fed to a trained deep learning model for inference to detect the function blocks from the bitstream. The deep learning model is trained using a dataset containing a lot of images transformed from various bitstreams. The process of bitstream-to-image transformation, dataset generation, deep learning training and inference could be implemented by scripts automatically. The rest of this section will describe each step in detail.

\subsection{Transformation from Bitstream to Image}
\label{3_2_sec_transformation_from_bitstream_to_image}

When transforming the bitstream file to the image, there are two challenges to address.

The first challenge is that the configuration bits for one element are not consecutive in a bitstream. The configuration memory in bitstream is arranged in frames, which are the smallest addressable segments of configuration memory space. For instance, the configuration bits for one CLB are distributed in several frames. In order to extract features from the image transformed from the bitstream, the image should reflect the adjacency between the elements in the programmable logic, which means the physical adjacent position relationship between the used elements.
The adjacency will be affected by the optimization algorithms using by the EDA tools when placing and routing. The implementations from different EDA tools or by different two times of the same function block have lots of common features, but not exactly the same.
The discontinuity of the configuration bits makes it difficult to reflect the adjacency between the elements in the programmable logic.

The second challenge is that not all configuration information in the bitstream file is useful for function block detection. There are two adverse effects that come with this fact: 1) Since our work focuses on the logic resources, the utilization information of other resources may confuse the function block detection. The programmable logic of FPGA includes CLBs, Input/Output Blocks (IOBs), Block RAMs (BRAMs) used for dense storage, Digital Signal Processors (DSPs) used for high-speed computing, and others. For instance, the utilization of BRAMs will vary with the array size of the function blocks. Accordingly, for the function blocks of the same kind, their BRAM utilization may vary. In contrast, the logic in the function block will not change according to the data size. As a result, those configuration bits, which are irrelevant to the logic resources and contribute nothing to function block detection, should be identified and ignored for the purpose of function detection. 2) The large size of images leads to a large image dataset and low speed of loading images during the deep learning training. For instance, if all of the configuration bits in a bitstream of Xilinx Zynq-7000 SoC ZC702 Evaluation Board (ZC702) are transformed into a three-channel color image, the image will have a dimension of 1280$\times$1080$\times$3, making it too large to be effectively handled by the detection model.

The above two challenges are addressed with the proposed bitstream to image transformation here. First and foremost, during such transformation, the bitstream format is analyzed and the mapping relationship between the CLB element and the configuration bits in the bitstream is found. Secondly, only configuration bits of CLBs are used for representation, which can compress effectively and drop useless information at the same time.

\begin{figure*}[!t]
\centerline{\includegraphics[width=7.1in]{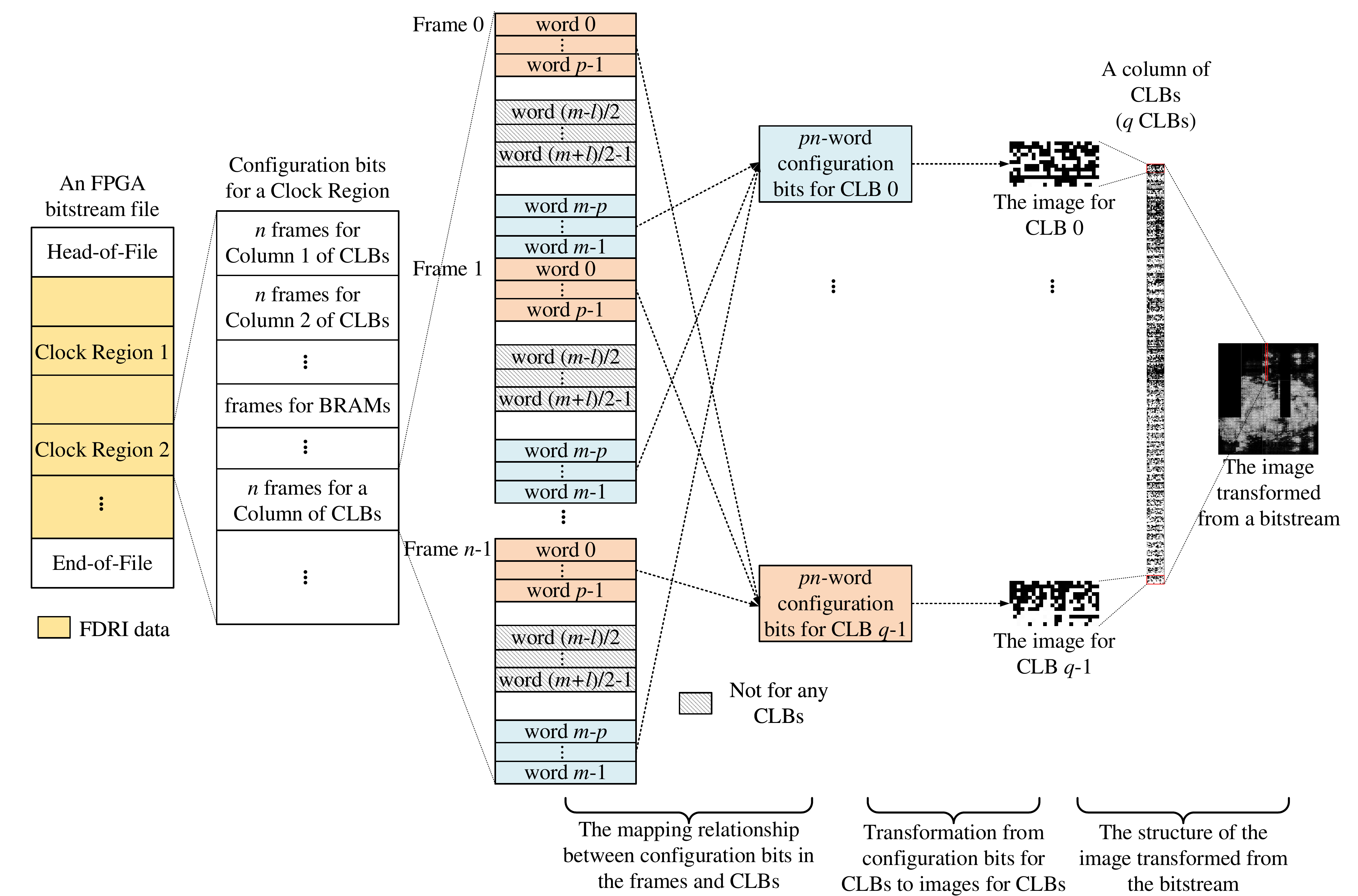}}
\caption{The transformation from an FPGA bitstream file to the image.}
\label{fig4}
\end{figure*}

\textbf{Mapping relationship between configuration bits and CLBs.}
An FPGA bitstream consists of three major parts:
Head-of-File, FDRI (Frame Data Register Input) data, and
End-of-File. Among these three parts, FDRI data contains
the configuration information for programmable logic. The programmable logic can be divided into several Clock Regions. As shown in Fig.~\ref{fig4}, each Clock Region consists of many columns of CLBs, and \emph{q} CLBs make up a column of CLBs. Each frame contains \emph{m} 32-bit words.
Since the open official document does not specify how CLBs are configured with the FDRI data, we have found that every successive \emph{n} frames of the FDRI data configure a column of CLBs (\emph{q} CLBs). Except for the \emph{l} words in the middle of the frame, every \emph{p} words in the left \emph{m}-\emph{l} words of a frame correspond to a CLB from the bottom to the top of the column.
For the two-slice CLBs, the \emph{p} words configure separately the two slices in a CLB from left to right. Therefore, each CLB in a column of CLBs needs \emph{p}$\times$\emph{n} words to configure, which are distributed at the same location in the \emph{n} frames.

\textbf{Transformation from configuration bits for CLBs to images for CLBs.}
According to the analysis of the bitstream format, each CLB is allocated 4\emph{p}$\times$\emph{n} bytes configuration memory from successive \emph{n} frames. For the two-slice CLBs, each slice is allocated \emph{s}=2\emph{p}$\times$\emph{n} bytes. Each slice in a one-slice CLB is allocated \emph{s}=4\emph{p}$\times$\emph{n} bytes.
Since the configuration bits are used to configure a separate slice, the bits with non-zero values indicate that this slice is used in this implementation. There are three methods to transform the \emph{s}-byte configuration data of a slice into a separate image with the proper height and width, which can indicate the utilization of the slice.

\begin{figure}[!t]
\centerline{\includegraphics[width=3.5in]{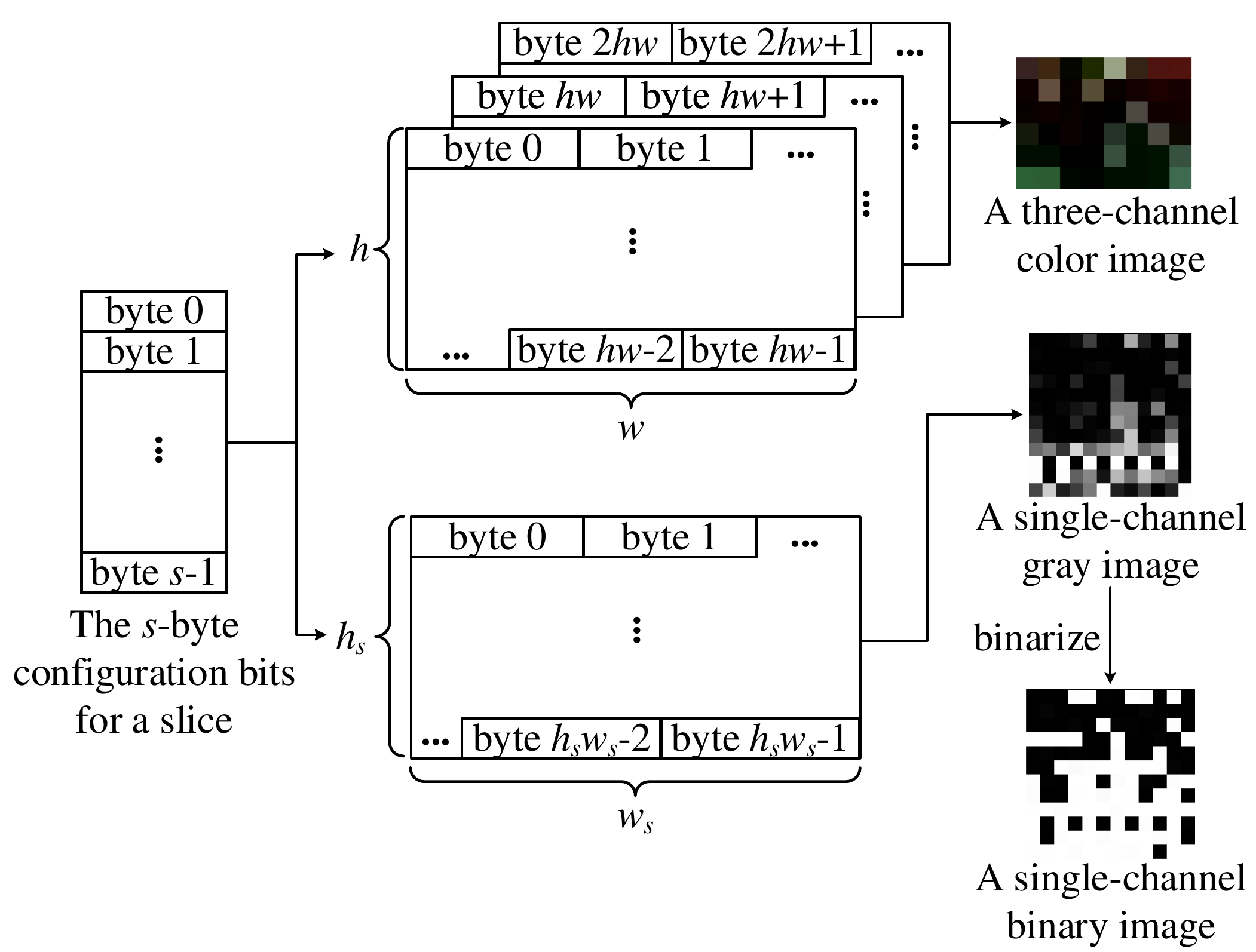}}
\caption{The \emph{s}-byte configuration bits for a slice are transformed into a three-channel color image with \emph{h}$\times$\emph{w} pixels, a single-channel gray image with \emph{$h_s$}$\times$\emph{$w_s$} pixels or a single-channel binary image with \emph{$h_s$}$\times$\emph{$w_s$} pixels.}
\label{fig6}
\end{figure}

The first method is to transform the data into a three-channel color image. As is shown in Fig.~\ref{fig6}, the configuration bits of a slice can be transformed into a three-channel color image with \emph{h}$\times$\emph{w} pixels. Each pixel, since they have three channels, needs three bytes. The configuration bits are considered as the image data arranged a three-tuple of (channel, height, width).

The second method is to transform the data into a single-channel gray image. Compared to the three-channel image, the data for one pixel in the single-channel image is continuous in the configuration data for a slice. The configuration bits of a slice can be transformed into a gray image with \emph{$h_s$}$\times$\emph{$w_s$} pixels. Each pixel requires one byte of storage.

The third method is to transform the data into a single-channel binary image. The single-channel binary image is obtained after the binarization of the single-channel gray image. The pixels with the non-zero values in the gray image are set one in the binary image. The binary image uses zeros and ones to represent whether the slices are used or not.

\textbf{Structure of the image transformed from bitstream.}
The image transformed from an FPGA bitstream is divided into \emph{a}$\times$\emph{b} blocks since each block in the several Clock Regions corresponds to a CLB. The parameter \emph{a} and \emph{b} are determined by the FPGA device, and are the same for the three methods of transformation from configuration bits to images for CLBs. As shown in Fig.~\ref{fig4}, the transformation to image for each CLB is done separately and the transformation results of all of the CLBs are aggregated to obtain the entire image. The adjacency between the CLBs in the programmable logic comes from the aggregation of the images for each CLB into the entire image.

\subsection{Deep Learning Model and Training}
\label{3_2_sec_DNN}

Since the deep learning techniques have not been applied to the FPGA function block detection, our work makes use of the transformation from bitstream to image and the image feature extraction capabilities of DNNs. YOLOv3~\cite{Redmon_YOLOv3_2018} and SSD~\cite{Liu_ECCV2016_SSD} are two classical one-stage deep learning-based object detection algorithms, which have fast speed and high accuracy at the same time.

\textbf{YOLOv3.} YOLOv3 consists of 75 convolution layers. 
The backbone of YOLOv3 is Darknet-53~\cite{Redmon_YOLOv3_2018}. There are three output convolution layers with different sizes of feature maps for detecting objects of different sizes. The three output convolution layers have the same filter number according to the class number. The input picture is divided into grids, and each grid cell predicts \emph{box\_number} bounding boxes (\emph{box\_number} is 3 for YOLOv3). One objectness prediction, \emph{C} class predictions for \emph{C} classes, and 4 box offsets are predicted for each bounding box. The objectness prediction quantifies how likely the image in the box contains a generic object~\cite{Alexe_TPAMI2012_objectness}. Thus, the filter number of the output layers is 3$\times$(1+\emph{C}+4). For example, we apply YOLOv3 to the detection of 10 kinds of function blocks (\emph{C}=10), so the filter number of the output layers should be 45. For the first convolution layer, the input channel number is determined by the channel number of the images fed into the deep neural network.

When training YOLOv3, we take the pre-trained weights for the COCO dataset~\cite{COCOdataset_ECCV2014} as initial weights. During the first 50 epochs of the training process, the front layers are frozen to get a stable loss value, except the last three output convolution layers. From the 51st to the 100th epoch, all of the layers are unfrozen and trained with a smaller learning rate. When the training is finished, choose the model with the smallest validation loss value as the final model. 

\textbf{SSD.} SSD consists of 29 convolution layers and 4 max-pooling layers. 
The backbone of SSD is VGG-16~\cite{Simonyan_ICLR2015_VGG}. There are six output convolution layers for objects of different sizes. Similar to YOLOv3, the filter number of the output layers is \emph{box\_number}$\times$(\emph{C}+4). In SSD, \emph{box\_number} can be 4 or 6 for different output layers. Thus, the filter number of the output layers is 56 or 84 when the class number \emph{C} is 10.

When training SSD, we load the trained weights of VGG-16 as initial weights for the front layers. In the first stage of training, the front layers are frozen with the weights from the VGG-16 model. Then, the whole network is trainable in the second stage of training. 

\textbf{Generation of dataset.}
The bitstreams are transformed into images, which are gathered into a dataset for deep learning training and testing. Each bitstream file implements an algorithm and each algorithm contains one or more function blocks. In a practical application, one kind of function block has different constructions, such as the original one with no special design and the pipelined one. Therefore, each kind of function block is implemented in one or two constructions when the bitstreams are generated.

Multiple bitstream files containing various kinds of function blocks are needed for the training of the deep network. We generate a large number of bitstreams by EDA toolset (Xilinx Vivado). The constraint of the implementation region makes the function blocks placed in different locations in different bitstreams. Tcl (Tool Command Language)~\cite{Xilinx_UG835_2018} of Xilinx Vivado is used instead of a graphical user interface (GUI). The categories and locations of the function blocks in the FPGA device diagram can be extracted from the EDA toolset when the bitstreams are generated by Tcl scripts. Finally, the bitstream files are transformed into images using the Python scripts. And the Python scripts process the label information into annotation files for deep learning at the same time.

\section{Experimental Results}
\label{4_sec_Experimental_results}

\subsection{Experimental Setup}\label{4_1_sec_Experimental_setup}

For evaluation purposes, we use Xilinx Zynq-7000 SoCs and Xilinx Zynq UltraScale+ MPSoCs to evaluate our proposed methodology. All of the experiments in this section are performed in the following experiment setup unless explicitly stated otherwise. All the bitstream files are generated without encryption by the Xilinx Vivado design suite, including Vivado 2016.3, Vivado 2017.2, Vivado 2017.4, Vivado 2018.3, and Vivado 2019.2. The scripts used for transforming bitstreams into images are running in Python 2.7.15. The training and testing of the deep learning are run in Keras 2.2.5 based on TensorFlow 1.10.0 for GPUs, with Python 3.5.6. A server running CentOS Linux 7.6, with an NVIDIA Tesla P100 GPU, is used to perform all of the experiments.

There are 10 kinds of function blocks to detect. The DNN YOLOv3 is used as the deep learning-based object detection algorithm unless otherwise stated. The DNN SSD is only used in the experiments in Section~\ref{4_5_6_sec}. The parameters for the training process of YOLOv3 and SSD are listed in TABLE~\ref{table1}.

\begin{table}[!t]
\caption{Parameters for the Training Process of YOLOv3 and SSD.}
\begin{center}
\begin{threeparttable}[b]
\begin{tabular}{|c|c|c|c|}
\hline
\multicolumn{2}{|c|}{Deep neural network} & YOLOv3  & SSD \\
\hline
\multicolumn{2}{|c|}{Input size$^{\mathrm{a}}$} & \tabincell{c}{416$\times$416$\times$1 or\\
416$\times$416$\times$3}
  & \tabincell{c}{300$\times$300$\times$1 or \\
  300$\times$300$\times$3} \\
\hline
\multicolumn{2}{|c|}{Optimizer} & Adam~\cite{Kingma_ICLR2015_adam}  & Adam \\
\hline
{The first stage } &	Batch size &	32 &	32 \\
\cline{2-4}
{of training} &	Learning rate &	0.001 & 0.001\\
\hline
{The second stage } &	Batch size &	16 &	32 \\
\cline{2-4}
{of training} &	Learning rate &	0.0001 & 0.001\\
\hline

\end{tabular}
\label{table1}
\begin{tablenotes}
  \item[a] The images are resized to the input size of DNN before fed into the DNN. The gray/binary images have 1 channel and the color images have 3 channels.
\end{tablenotes}
\end{threeparttable}
\end{center}
\end{table}

To characterize the performance of the object detector quantitatively, mAP under specific IoU is used as the performance metric, which takes account of both precision and recall. In general, the performance is good when the IoU for the detected box and the ground truth is more than 0.5. In our work, two metrics are used, imitating the COCO dataset. One is the mAP at IoU=0.5 (mAP@0.5), which is the metric for the PASCAL VOC dataset~\cite{Everingham_2010_VOCdataset}. The other one is the mAP at IoU=0.75 (mAP@0.75), which is stricter than mAP@0.5.

\subsection{Bitstream Format Analysis}
\label{4_2_sec_bitstream_analysis}

For the purpose of finding the mapping relationship used for the transformation from bitstream to image, this work analyzes the bitstream format of Xilinx Zynq-7000 SoCs and Xilinx Zynq UltraScale+ MPSoCs.
For Xilinx Zynq-7000 SoCs, a CLB element contains two slices, and each slice consists of 4 LUTs and 8 Flip-Flops (FFs)~\cite{Xilinx_UG585_2018}. For Xilinx Zynq UltraScale+ MPSoCs, a CLB element contains one slice, and each slice consists of 8 LUTs and 16 FFs~\cite{Xilinx_UG574_2017}. There are two types of slices, SLICEL (logic) and SLICEM (memory) respectively.
The specific bitstream format information can be found in the open official documents~\cite{Xilinx_UG470_2018,Xilinx_UG570_2020}.

\begin{figure}[!t]
\centerline{\includegraphics[width=3.5in]{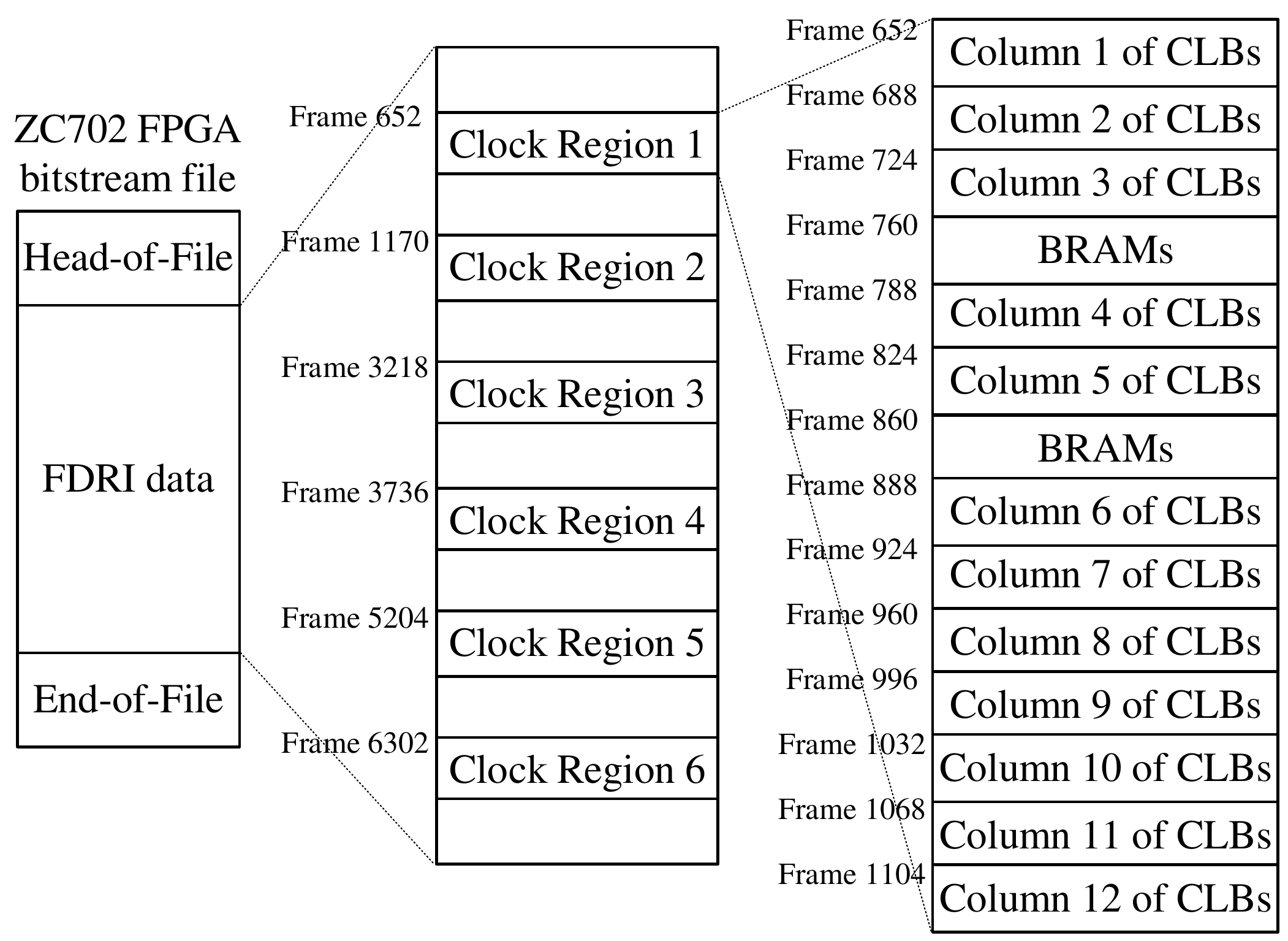}}
\caption{The positions of some first frames in the bitstream of ZC702 FPGA are shown. The number of frames configuring a column of CLBs (\emph{n}) is 36 for ZC702 FPGA.}
\label{fig5}
\end{figure}

The positions of every first frame are necessary when transforming the bitstreams into images, which can be found by configuring the different columns of CLBs repeatedly.
As an example, the positions of some first frames in the bitstream of ZC702 FPGA are shown in Fig.~\ref{fig5}. The number of frames configuring a column of CLBs (\emph{n}) is 36 for ZC702 FPGA, and a column of BRAMs need 28 frames to configure. 

\subsection{Transformation Result from Bitstream}
\label{4_3_sec_representation}

According to the bitstream format analysis results in Section~\ref{4_2_sec_bitstream_analysis}, the bitstreams are transformed into images. Taking Xilinx Zynq-7000 SoC Z-7020, Xilinx Zynq-7000 SoC Z-7030, and Xilinx Zynq UltraScale+ MPSoC ZU9EG as examples, the parameters for transforming the bitstreams into images are listed in TABLE~\ref{table4}.

\begin{table}[!t]
\caption{Parameters for Transforming the Bitstreams into Images, taking Xilinx Zynq-7000 SoC Z-7020, Xilinx Zynq-7000 SoC Z-7030, and Xilinx Zynq UltraScale+ MPSoC ZU9EG as Examples.}
\begin{center}
\begin{threeparttable}[b]
\begin{tabular}{|c|c|c|c|}
\hline

Device name &	\tabincell{c}{Xilinx \\Zynq-7000 \\SoC Z-7020$^{\mathrm{a}}$}&	\tabincell{c}{Xilinx \\Zynq-7000 \\SoC Z-7030$^{\mathrm{b}}$}&	\tabincell{c}{Xilinx Zynq\\ UltraScale+ \\MPSoC \\ZU9EG$^{\mathrm{c}}$}\\
\hline

\tabincell{c}{Number of \\blocks the \\device diagram \\divided (\emph{a}$\times$\emph{b})}&	150$\times$57&	200$\times$60&	\tabincell{c}{420$\times$97\\ (46 for SLICEL,\\ 51 for SLICEM)}\\
\hline

\tabincell{c}{Bitstream \\length (bits)}&	32,364,512	& 47,839,328	&212,086,240\\
\hline


\tabincell{c}{Size of \\color image \\of a slice}&	6$\times$8$\times$3&	6$\times$8$\times$3	&\tabincell{c}{7$\times$9$\times$3 \\for SLICEL,\\7$\times$23$\times$3 \\for SLICEM}\\
\hline

\tabincell{c}{Size of the \\entire color\\ image (bytes)}&	900$\times$912$\times$3	&1200$\times$960$\times$3&	2940$\times$1587$\times$3\\
\hline

\tabincell{c}{Compression \\ratio of color \\ image(\%)}	&60.87	&57.79&	52.80\\
\hline

\tabincell{c}{Size of the \\gray/binary\\ image of a slice}&	12$\times$12$\times$1&	12$\times$12$\times$1	&\tabincell{c}{12$\times$15$\times$1 \\for SLICEL,\\12$\times$41$\times$1 \\for SLICEM}\\
\hline

\tabincell{c}{Size of the \\entire gray/\\binary image\\ (bytes)}&	1800$\times$1368$\times$1	&2400$\times$1440$\times$1&	5040$\times$2781$\times$1\\
\hline

\tabincell{c}{Compression \\ratio of \\gray/binary\\  image(\%)}	&60.87	&57.79&	52.87\\
\hline

\end{tabular}
\label{table4}
\begin{tablenotes}
    \item[a] Evaluated on Xilinx Zynq-7000 SoC ZC702 Evaluation Board.
    \item[b] Evaluated on Xilinx xc7z030fbg484-3 FPGA.
    \item[c] Evaluated on Xilinx Zynq UltraScale+ ZCU102 Evaluation Board.
\end{tablenotes}
\end{threeparttable}
\end{center}

\end{table}




Note that the bitstream length of ZC702 FPGA is 3.86 MiB. The three-channel color image transformed from bitstream has a size of 900$\times$912$\times$3 bytes. So the transformation compresses effectively the color image to
(900$\times$912$\times$3$\times$8 bits/32,364,512 bits)$\times$100\%=60.87\%
of the original bitstream data. The effective compression mainly benefits from the drop of the configuration bits irrelevant to the logic resources. The parameters for transforming the bitstreams into images are set differently for the different FPGA devices and determined by the bitstream format information of FPGA devices.

\begin{figure}[!t]
\centering
\subfigure[]{
\includegraphics[width=1.58in]{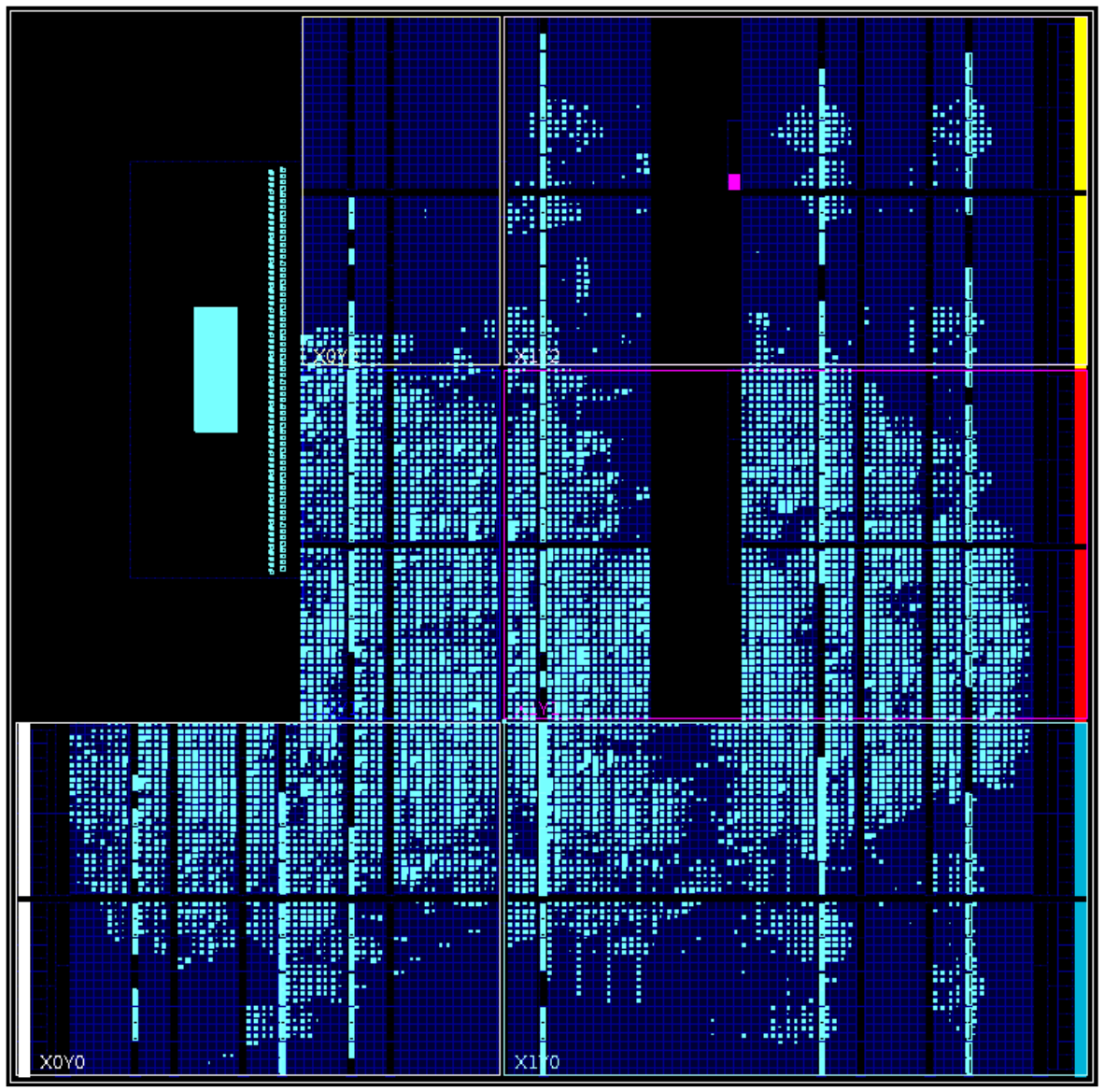}
}
\hfil
\subfigure[]{
\includegraphics[height=1.56in,width=1.58in]{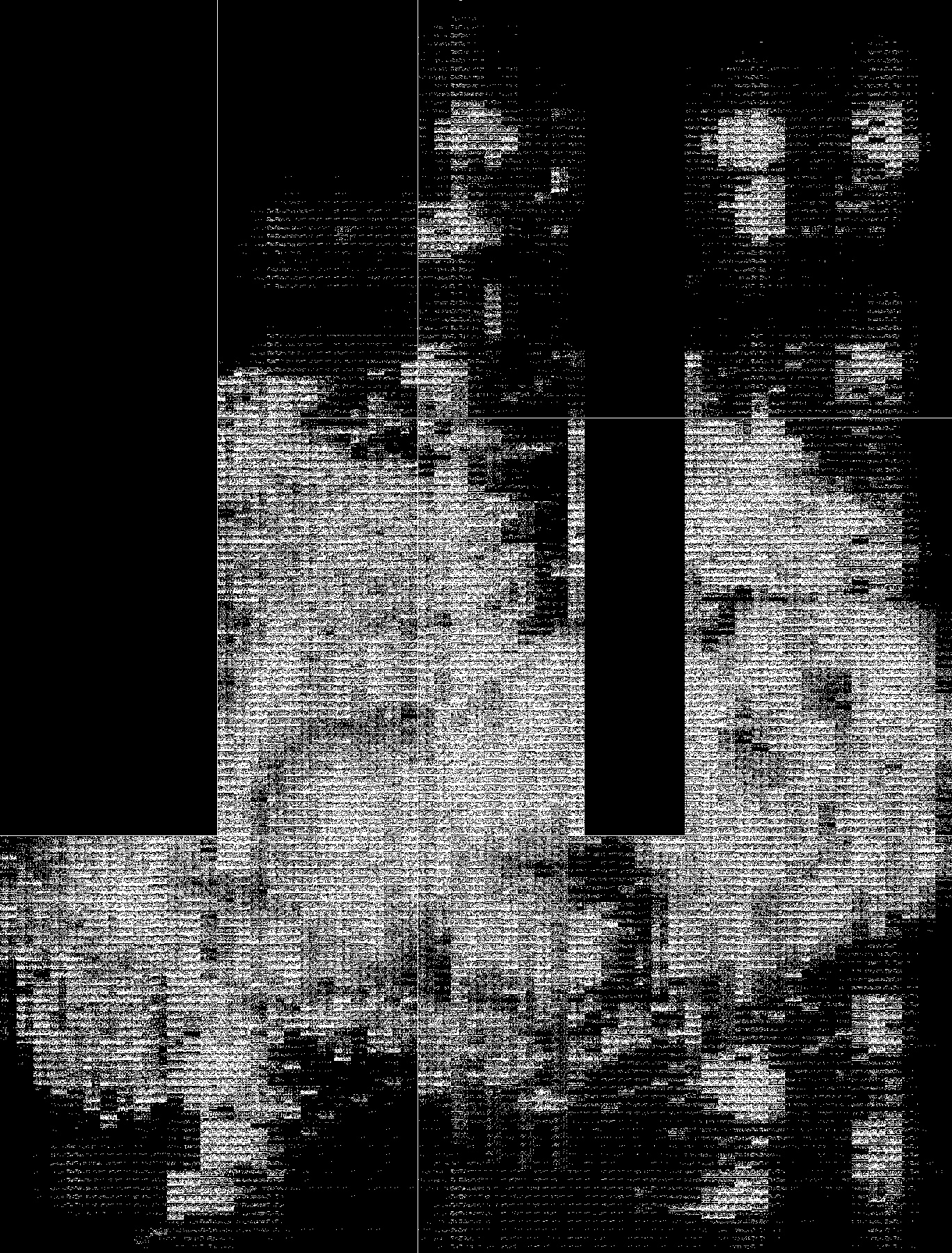}
}
\caption{Vivado implemented designs and the images transformed from bitstreams. (a) Vivado implemented design and (b) the single-channel binary image transformed from bitstream of Xilinx Zynq-7000 SoC Z-7020. }
\label{fig7}
\end{figure}

The Vivado implemented design and the image transformed from the bitstream of Xilinx Zynq-7000 SoC Z-7020 are shown in Fig.~\ref{fig7}. The dark blue area in the Vivado implemented design represents columns of CLBs, the configuration bits of which are used for the bitstream representation. The corresponding relationship between the Vivado implemented design and the image transformed from the bitstream for the same FPGA device proves that the mapping relationship between configuration bits and CLBs found by this work is correct and the image transformed from bitstream can reflect the adjacency of the programmable logic.

In summary, the two challenges mentioned in Section~\ref{3_2_sec_transformation_from_bitstream_to_image} have been overcome by the proposed transformation from bitstream to image.

\subsection{Dataset Description}
\label{4_4_sec_dataset}

For the purpose of training and testing the DNN models, a large number of bitstreams, which implement FPGA designs on ZC702 FPGA, are generated to make up the dataset. There are 15 kinds of application-specific encryption algorithms chosen for generating 18,286 bitstream files and these encryption algorithms contain 10 kinds of cryptographic operators. The application-specific encryption algorithms used for generating the dataset and the cryptographic operators contained are listed in TABLE~\ref{table5}. Each encryption algorithm used in this work contains up to 3 kinds of cryptographic operators. Each kind of cryptographic operator is implemented in one or two constructions. The pipeline means the cryptographic operator is implemented in a pipeline design, and the module means the cryptographic operator is implemented without special designs. For example, a bitstream implementing the encryption algorithm used for NTLM (NT LAN Manager) contains a Message Digest Algorithm 4 (MD4) pipeline or an MD4 module. A bitstream implementing the encryption algorithm used for PDF-R2 contains an MD5 pipeline and an RC4 module.

\begin{table*}[!t]
\caption{Components of the Training Set and Test Set.}
\begin{center}
\begin{tabular}{|c|c|c|c|c|c|}
\hline
\tabincell{c}{Vivado\\ version}	& \tabincell{c}{Applications of \\encryption \\algorithms}&\tabincell{c}{Number of \\cryptographic\\ operators\\ contained}&\tabincell{c}{Cryptographic\\ operators}	& \tabincell{c}{Implementation constructions}&\tabincell{c}{Number of\\ bitstreams}\\
\hline
\hline
&	\emph{\tabincell{c}{For training and testing}}	&&&&\\
\hline
\multirow{13}{*}{2017.4}
&LM &1	&DES	&\tabincell{c}{DES pipeline; DES module}	&1,294\\

&NTLM	&1	&MD4	&\tabincell{c}{MD4 pipeline; MD4 module}	&1,135\\

&PDF-R2	&2	&MD5, RC4	&\tabincell{c}{MD5 pipeline, RC4 module}&2,252\\

&OFFICE 2003	&2	&MD5, RC4	&\tabincell{c}{MD5 pipeline, RC4 module}	& 419\\

&TrueCrypt	&3	&AES, Serpent, Twofish	&\tabincell{c}{AES module, Serpent module, Twofish module}	&1,842\\

&OFFICE 2007	&1	&SHA-1	&SHA-1 pipeline	&429\\

&WPA-2	&1	&SHA-1	&SHA-1 pipeline	&451\\

&	WINZIP	&1	&SHA-1	&SHA-1 pipeline	&542\\

&RAR3	&1	&SHA-1	&SHA-1 module	&165\\

&RAR5	&1	&SHA-256	&SHA-256 pipeline	&245\\

&7ZIP	&1	&SHA-256	&SHA-256 pipeline	&242\\

&OFFICE 2013	&1	&SHA-512	&SHA-512 module &303\\

&LINUX-SHA512	&1	&SHA-512	&SHA-512 module	&728\\
\hline
&Total				&&&&10,047\\
\hline
\hline
&	\emph{For testing only}&&&&	\\		
\hline	
\multirow{2}{*}{2016.3}
&	PDF-R2&	2&	MD5, RC4	&MD5 pipeline, RC4 module	&587\\
&	PDF-R5&	2&	MD5, RC4&	MD5 pipeline, RC4 module&	1,107\\
\hline
\multirow{2}{*}{2017.2}
&	PDF-R2&	2	&MD5, RC4&	MD5 pipeline, RC4 module&	587\\
&PDF-R5&	2	&MD5, RC4&	MD5 pipeline, RC4 module&	1,407\\
\hline
\multirow{2}{*}{2017.4}	
&PDF-R5	&2&	MD5, RC4	&MD5 pipeline, RC4 module&	1,107\\
&	OFFICE 2010&	1	&SHA-1	&SHA-1 pipeline	&232\\
\hline
\multirow{2}{*}{2018.3}
&	PDF-R2&	2	&MD5, RC4&	MD5 pipeline, RC4 module&	587\\
&PDF-R5&	2	&MD5, RC4&	MD5 pipeline, RC4 module&	1,144\\
\hline
\multirow{2}{*}{2019.2}
&	PDF-R2&	2	&MD5, RC4&	MD5 pipeline, RC4 module&	587\\
&PDF-R5&	2	&MD5, RC4&	MD5 pipeline, RC4 module&	894\\
\hline
&	Total		&&&&		8,239\\
\hline

\end{tabular}
\label{table5}
\end{center}
\end{table*}

In order to arrange the experiments reasonably, 13 kinds of encryption algorithms in TABLE~\ref{table5} are chosen to make up the training set and the test set, including 10,047 bitstreams generated by Xilinx Vivado 2017.4 totally. The bitstreams are divided into the training set and the test set randomly by 4:1. Two kinds of encryption algorithms, used for PDF-R5 and OFFICE 2010, implemented by Xilinx Vivado 2017.4 are just used for testing. Besides, to explore the effect of EDA tools, the bitstreams generated by Xilinx Vivado 2016.3, Vivado 2017.2, Vivado 2018.3, and Vivado 2019.2 are used to test the performance of the trained model. These bitstreams are transformed into images to make up the dataset for training and testing in Section~\ref{4_5_sec_detection}.

\subsection{Function Block Detection}
\label{4_5_sec_detection}


\subsubsection{Evaluation results on different transformation from bitstream to image}\label{4_5_0_sec}

The first 13 kinds of encryption algorithms listed in TABLE~\ref{table5} are chosen to make up the training set and the test set. Since the images transformed from the bitstreams of the 13 kinds of encryption algorithms are divided into the training set and the test set randomly by 4:1, the test set has the same distribution as the training set.

In this experiment, there are five models trained in five different datasets, which are transformed from the same bitstreams into different image representation, namely 1) three-channel color image, 2) pseudo color image from the single-channel gray image, 3) pseudo color image from the single-channel binary image, 4) single-channel gray image, and 5) single-channel binary image. The pseudo color image with three channels is transformed from the single-channel image. Since the input channel number of the pre-trained model is three, using the pseudo color image makes it unnecessary to train the first convolution layer from random initialization. For the single-channel image, the input channel number is one.

We test the five models on the test set with the same distribution as the training set. The evaluation results are listed in TABLE~\ref{table_4_5_0}. It is demonstrated that the model trained in the single-channel binary images has the best accuracy. The models trained in the pseudo color images have a worse accuracy than three-channel color images or single-channel images.

\begin{table}[!t]
\caption{Evaluation Results of Five Models with Different Transformation from Bitstream to Image on the Test Set.}
\begin{center}
\footnotesize
\begin{tabular}{|c|c|c|}
\hline
Image representation 	&mAP@0.5 (\%)&	mAP@0.75 (\%)\\
\hline
\tabincell{c}{ 1) three-channel color image}&	97.76	&95.57\\
\hline
 \tabincell{c}{2) pseudo color image from\\ the single-channel gray image }&	97.70&	95.16\\
\hline
\tabincell{c}{ 3) pseudo color image from \\the single-channel binary image } &	96.94	&94.81\\
\hline
 4) single-channel gray image &	98.02&	95.74\\
\hline
 5) single-channel binary image&	98.11	&95.79\\
\hline

\end{tabular}
\label{table_4_5_0}
\end{center}
\end{table}

The analyses of the evaluation results are as follows: 1) The evaluation results of the five models are close, since the five image representation of bitstream all can reflect the adjacency between the elements in the programmable logic. 2) The pseudo color images are transformed from the single-channel images, not directly from the data from the bitstream, which may lead to the worse accuracy. 3) The data for one pixel in the single-channel image is continuous in the configuration data for a slice, which is the reason for the better accuracy of the single-channel image representation than the three-channel color image representation. However, the data for the three channels of one pixel in the RGB color image is not continuous, and the data from different channels may disturb each other. 4) The single-channel binary image representation has a better accuracy than the single-channel gray image, because the binary image focuses the features of the utilization and the adjacency between the elements in the programmable logic. Actually what we care about is whether the slice is used in the implementation instead of the values of pixels. The binary image uses zeros and ones to represent whether the slices are used or not. The single-channel binary image representation is used in the following experiments.

\subsubsection{Evaluation results on the test set}
\label{4_5_1_sec} 
\begin{figure}[!t]
\centerline{\includegraphics[width=1.8in,height=1.78in]{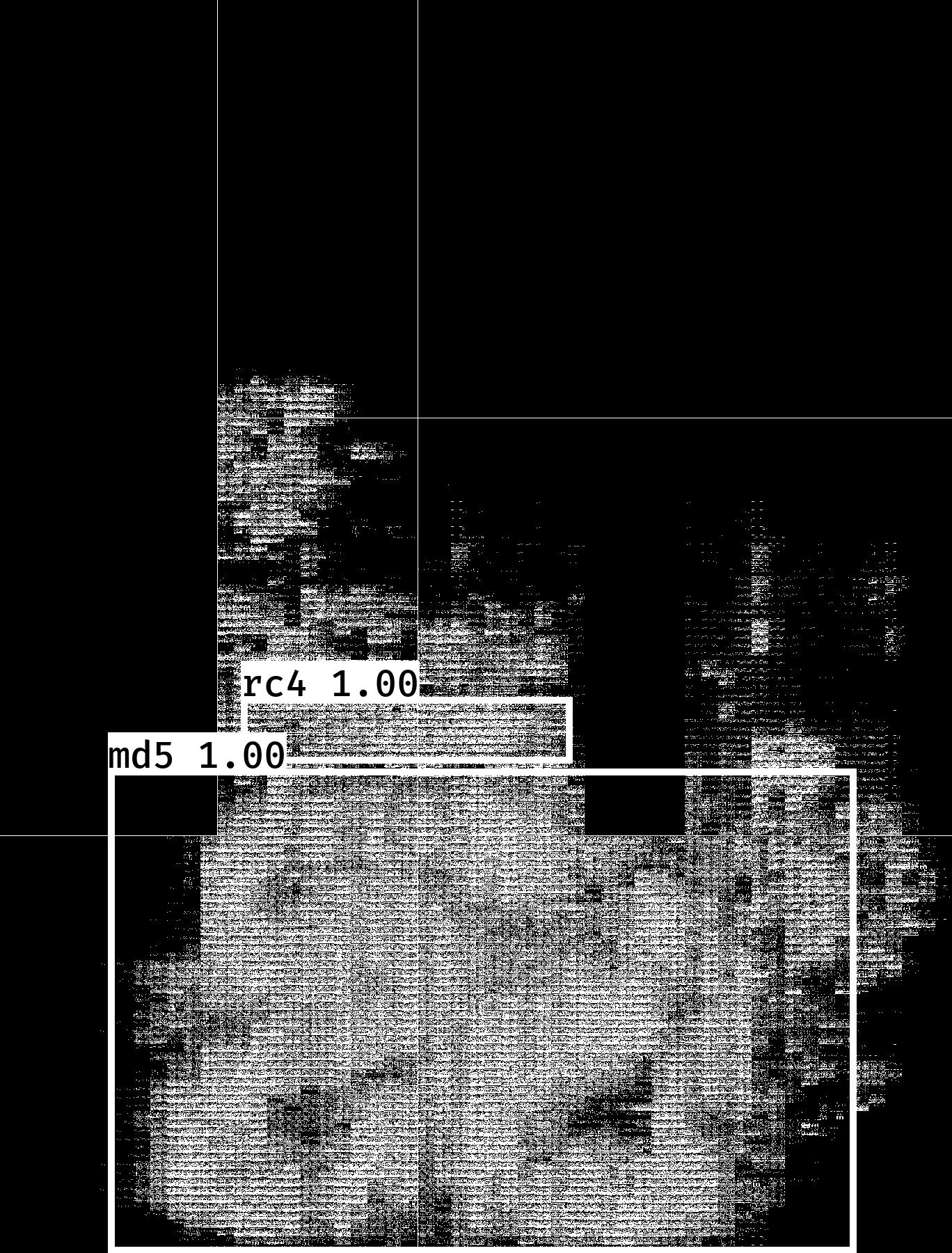}}
\caption{The function block detection result of a bitstream file implementing the encryption algorithm used for PDF-R2 on ZC702 FPGA.}
\label{fig8}
\end{figure}

In this experiment,
the training process is the same as Section~\ref{4_5_0_sec}. The evaluation result of the single-channel binary image representation is shown in detail in this section. The function block detection result of a bitstream file, which implements the encryption algorithm used for PDF-R2 on ZC702 FPGA, is shown in Fig.~\ref{fig8}, as an example. The function blocks included in this image are marked with boxes. Each box is labeled with the category and a classification probability.

TABLE~\ref{table6} shows the evaluation results evaluated quantitatively on the test set. AP (Average Precision) @0.5 of 10 kinds of cryptographic operators are all beyond 86.13\% and mAP@0.5 reaches 98.11\%. Even under the stricter metric, mAP@0.75 reaches 95.79\%. It is evident that the detector has a good detection performance on the test set with the same distribution as the training set.

\begin{table*}[!t]
\caption{Evaluation Results on the Test Set with the Same Distribution as the Training Set.}
\begin{center}
\begin{tabular}{|c|c|c|c|c|c|c|c|c|c|c|c|}
\hline
\tabincell{c}{Function blocks}&	\textbf{mAP}&	AES	&DES&	MD4&	MD5	&RC4	&Serpent&	SHA-1	&SHA-
256&	SHA-
512	&Twofish\\
\hline
AP@0.5 (\%)&	\textbf{98.11}&	94.97	&100.00&	100.00	&100.00	&99.95	&86.13	&100.00	&100.00	&100.00	&100.00\\
\hline
AP@0.75 (\%)&	\textbf{95.79}	&79.63	&99.88	&99.92&	98.34&	96.17&	84.59&	99.92 &100.00&	99.48&	100.00\\
\hline
\end{tabular}
\label{table6}
\end{center}
\end{table*}

The image transformed from bitstream can reflect the resource utilization of function blocks and the adjacency between the CLBs used. Different kinds of function blocks are different in these two aspects. The Secure Hashing Algorithm 256 (SHA-256) module nearly takes over all of the LUT resources of ZC702 FPGA. However, the RC4 module only occupies no more than 1\% of the FF resources and approximately 3\% of the LUT resources. When implemented in the same construction, two function blocks of the same kind are similar but appear in different locations in different images. Since the image transformed from bitstream keeps some characteristics of function blocks that can distinguish one kind of function block from another one, the proposed transformation from bitstream to image is proved effective and applicable for generating a dataset for deep learning. It also demonstrates that YOLOv3 learns successfully how to detect the function blocks from the images.

Some kinds of cryptographic operators are well detected, such as SHA-1 and SHA-256. It is because these kinds of cryptographic operators occupy a large area of the images. It is not difficult to detect a box with an IoU over 0.5 or 0.75 with the ground truth.
Some kinds of cryptographic operators have relatively low AP, such as Advanced Encryption Standard (AES) and Serpent. The reasons are as follows: 1) These kinds of cryptographic operators always occupy a small number of resources of FPGA. 2) It is more difficult to detect a function block from a system with various function blocks than from a system with a single function block. 

\subsubsection{Effectiveness of detecting cryptographic operators}\label{4_5_2_sec}

In this experiment, the trained model is tested on the bitstream files implementing the encryption algorithms, which do not appear in the training set, to confirm the capability to detect cryptographic operators. The training process is the same as mentioned in Section~\ref{4_5_0_sec}. We choose 1,339 bitstreams for testing in this experiment, which implement two kinds of encryption algorithms used for PDF-R5 and OFFICE 2010 and are generated by Vivado 2017.4. The function blocks in these bitstreams have the same implementation constructions as the ones in the training set.

TABLE~\ref{table7} shows the results of this experiment. The evaluation results on the encryption algorithms used for PDF-R5 and OFFICE 2010 show that the detector also has a good performance on the encryption algorithms not appearing in the training set, because the cryptographic operators in these encryption algorithms have the same implementation constructions as the ones in the training set. Although the RC4 module has appeared in the training set, the small occupied area of the RC4 module accounts for the relatively low AP. TABLE~\ref{table6} and TABLE~\ref{table7} demonstrate that the detector has the capability to detect the function blocks with the same constructions in the training set, no matter whether the encryption algorithms appear in the training set or not.

\begin{table}[!t]
\caption{Evaluation Results on the Encryption Algorithms not Appearing in the Training Set.}
\begin{center}
\begin{tabular}{|c|c|c|c|}
\hline
\multicolumn{2}{|c|}{Function blocks} & AP@0.5 (\%) &	AP@0.75 (\%)\\
\hline
\multirow{2}{*}{PDF-R5}
&	(1) MD5 pipeline&	100.00&	99.86\\
\cline{2-4}
&	(2) RC4 module&	96.37	&92.57\\
\hline
OFFICE 2010	&(3) SHA-1 pipeline&	100.00&	99.98\\
\hline

\end{tabular}
\label{table7}
\end{center}
\end{table}


\subsubsection{Effect of EDA tools}\label{4_5_3_sec}
In this experiment, the model is trained on the bitstreams generated by Vivado 2017.4 and tested on the bitstreams generated by the other versions of Xilinx Vivado, namely Vivado 2016.3, Vivado 2017.2, Vivado 2018.3, and Vivado 2019.2. The training process is the same as mentioned in Section~\ref{4_5_0_sec}. The bitstreams for testing implementing encryption algorithms used for PDF-R2 and PDF-R5, contain the function blocks with the same implementation constructions as the ones in the training set. The encryption algorithm used for PDF-R2 is included in the training set. However, the encryption algorithm used for PDF-R5 is not included. This experiment is set up to explore the effect of the EDA tools. Although the EDA tools are provided by the FPGA vendors, the corresponding EDA tools are updated continually.

\begin{figure}[!t]
\centering
\subfigure[]{
\includegraphics[width=3in]{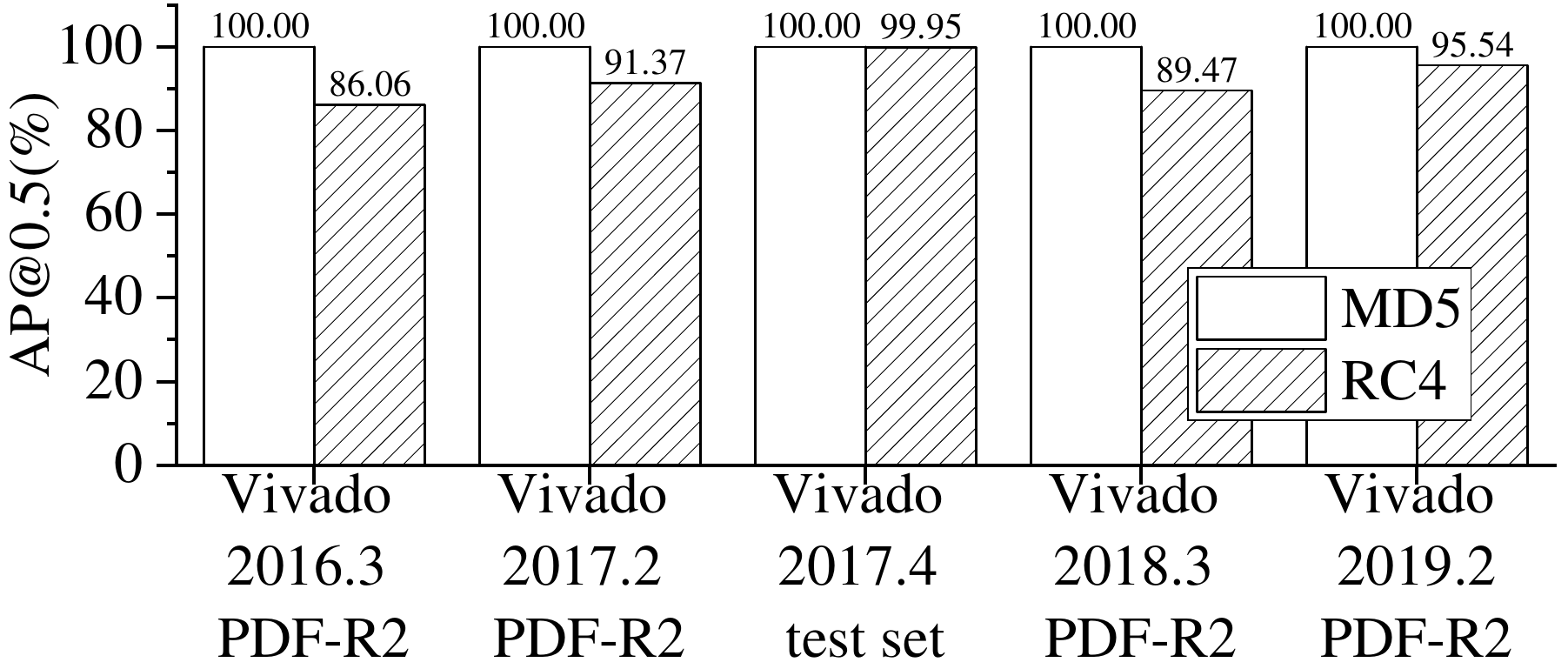}
}
\hfil
\subfigure[]{
\includegraphics[width=3in]{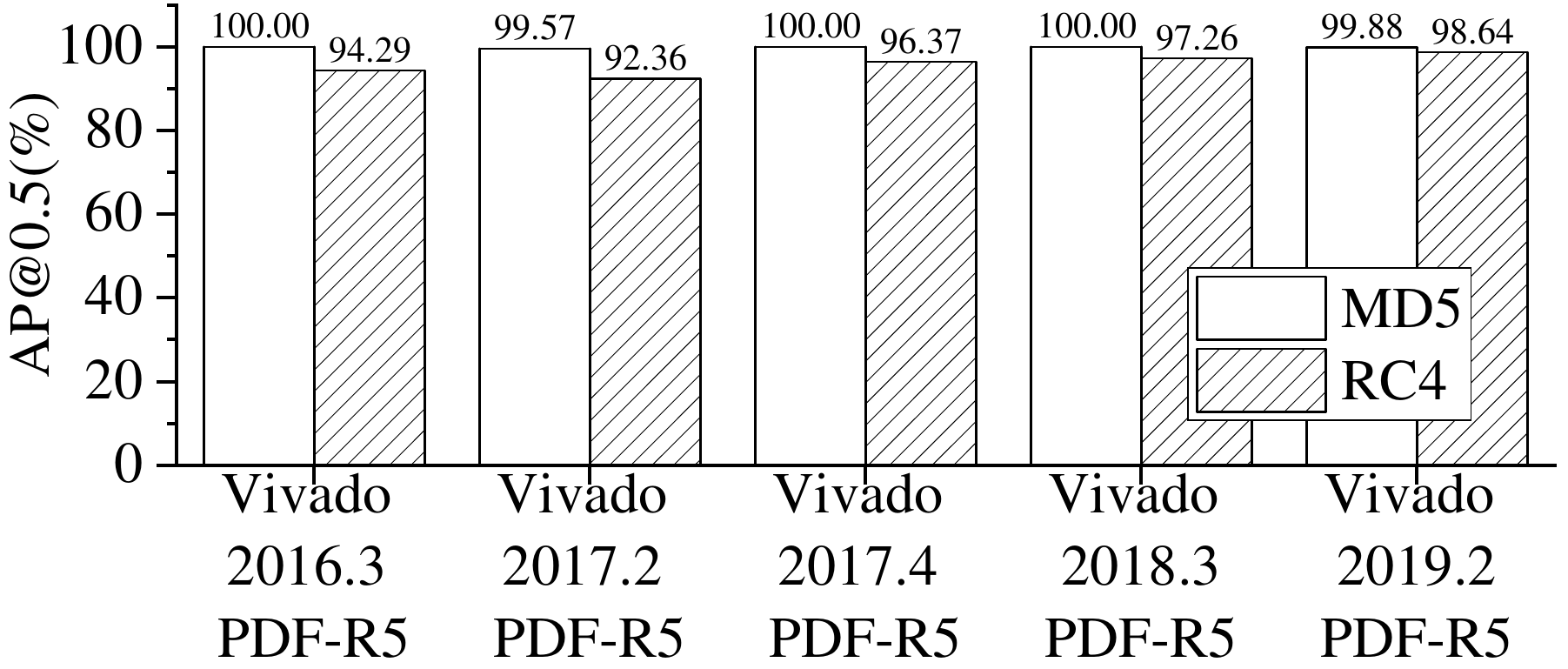}
}
\caption{Evaluation results on the bitstreams generated by different EDA tools. The AP@0.5 of MD5 and RC4 evaluated on the bitstreams of encryption algorithms used for (a) PDF-R2 and (b) PDF-R5 generated by other EDA tools, compared with the evaluation results on the bitstreams generated by Vivado 2017.4. }
\label{fig9}
\end{figure}

The evaluation results are shown in Fig.~\ref{fig9}. The evaluation results on bitstreams generated by other versions of Vivado are compared with the evaluation results on bitstreams generated by Vivado 2017.4. 
Although there are some differences in placing and routing when optimizing the same design on the different versions of Vivado toolset, there is no big difference in the resource utilization of function blocks and the effect of EDA tools on the detection accuracy is really slight. It is evident that the model trained on bitstreams generated by Vivado 2017.4 can detect the function blocks from the bitstreams generated by the other versions of Xilinx Vivado, although the detection accuracy of some function blocks, such as RC4, may decrease due to the really low resource utilization.

\subsubsection{Application of other deep learning-based object algorithms to function block detection}\label{4_5_6_sec}

In this experiment, SSD is applied to bitstream function block detection to demonstrate the effectiveness of the methodology. In the methodology, it is not necessary to choose YOLOv3 as the deep learning-based object detection algorithm. The training set and test set are the same as used in Section~\ref{4_5_1_sec}. The bitstreams are generated by Vivado 2017.4.

The evaluation results of SSD on the test set are listed in TABLE~\ref{table10}, together with the evaluation results of YOLOv3 in Section~\ref{4_5_1_sec}. The evaluation results of SSD show that SSD also has the capability to detect function blocks. The methodology has generality to some degree and the other deep learning-based object detection algorithms with high performance can be applied. Compared with the performance of YOLOv3, the performance of SSD is evidently lower than YOLOv3.
However, mAP@0.75 of SSD is much lower than YOLOv3. It is shown that the location accuracy of SSD for function blocks is poorer than YOLOv3 in this scenario.

\begin{table}[!t]
\caption{Evaluation Results of YOLOv3 and SSD on the Test Set with the Same Distribution as the Training Set.}
\begin{center}
\begin{tabular}{|c|c|c|}
\hline

\tabincell{c}{Deep neural network \\(Input size)}&	mAP@0.5 (\%)&	mAP@0.75 (\%)\\
\hline
\tabincell{c}{YOLOv3 (416$\times$416$\times$1)}	&98.11 &	95.79\\
\hline
\tabincell{c}{  SSD (300$\times$300$\times$1)	}&90.80&	68.05\\
\hline

\end{tabular}
\label{table10}
\end{center}
\end{table}

\subsection{Processing Time}
\label{4_6_secProcessing_time}
\begin{figure}[!t]
\centerline{\includegraphics[width=3in]{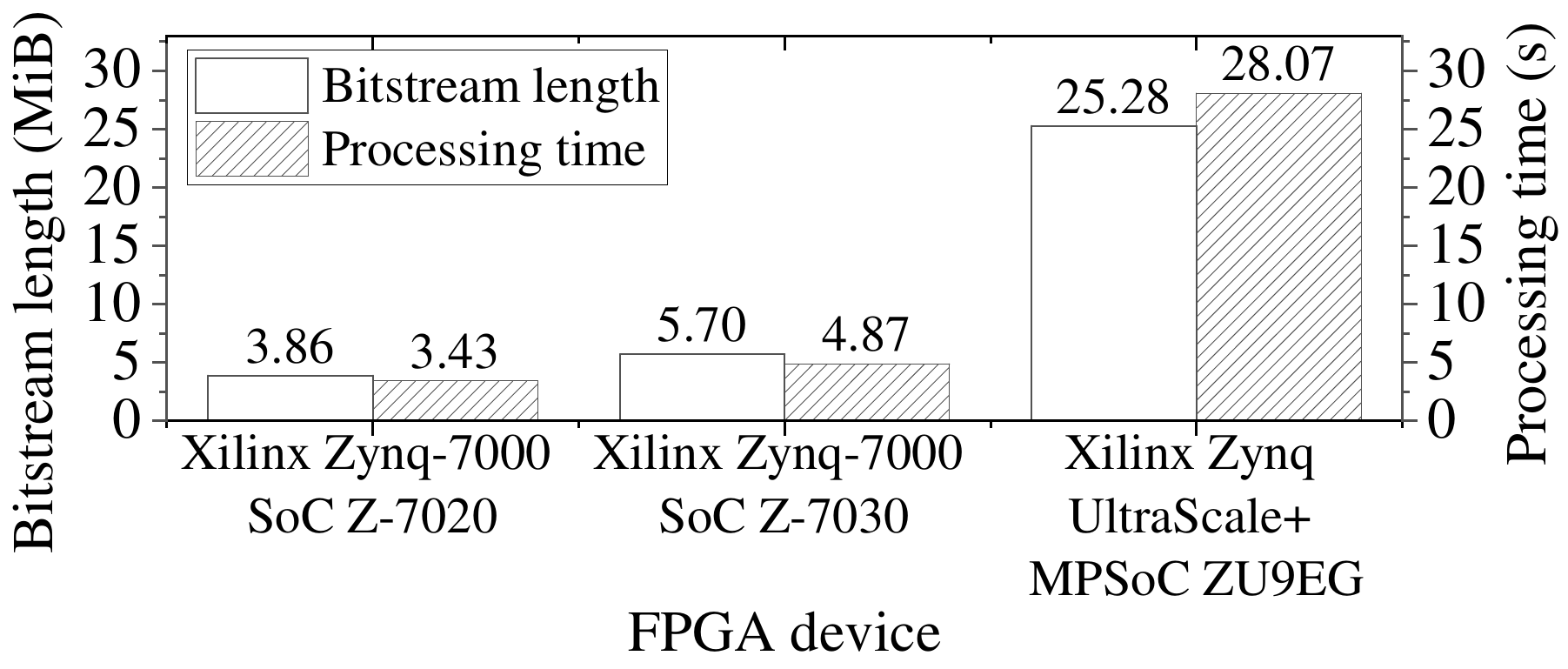}}
\caption{The processing time of transforming a bitstream into an image on a single Intel Xeon Gold 5118 CPU@2.30GHz.}
\label{fig11}
\end{figure}

Taking Xilinx Zynq-7000 SoC Z-7020, Xilinx Zynq-7000 SoC Z-7030, and Xilinx Zynq UltraScale+ MPSoC ZU9EG as examples, the processing time of transforming a bitstream into an image is reported in Fig.~\ref{fig11}, together with the bitstream length. The processing time is measured on a single Intel Xeon Gold 5118 CPU@2.30GHz. It is evident that the processing time of transforming a bitstream into an image is almost proportional to the bitstream length, which is determined by the FPGA device.

\begin{figure}[!t]
\centerline{\includegraphics[width=3in]{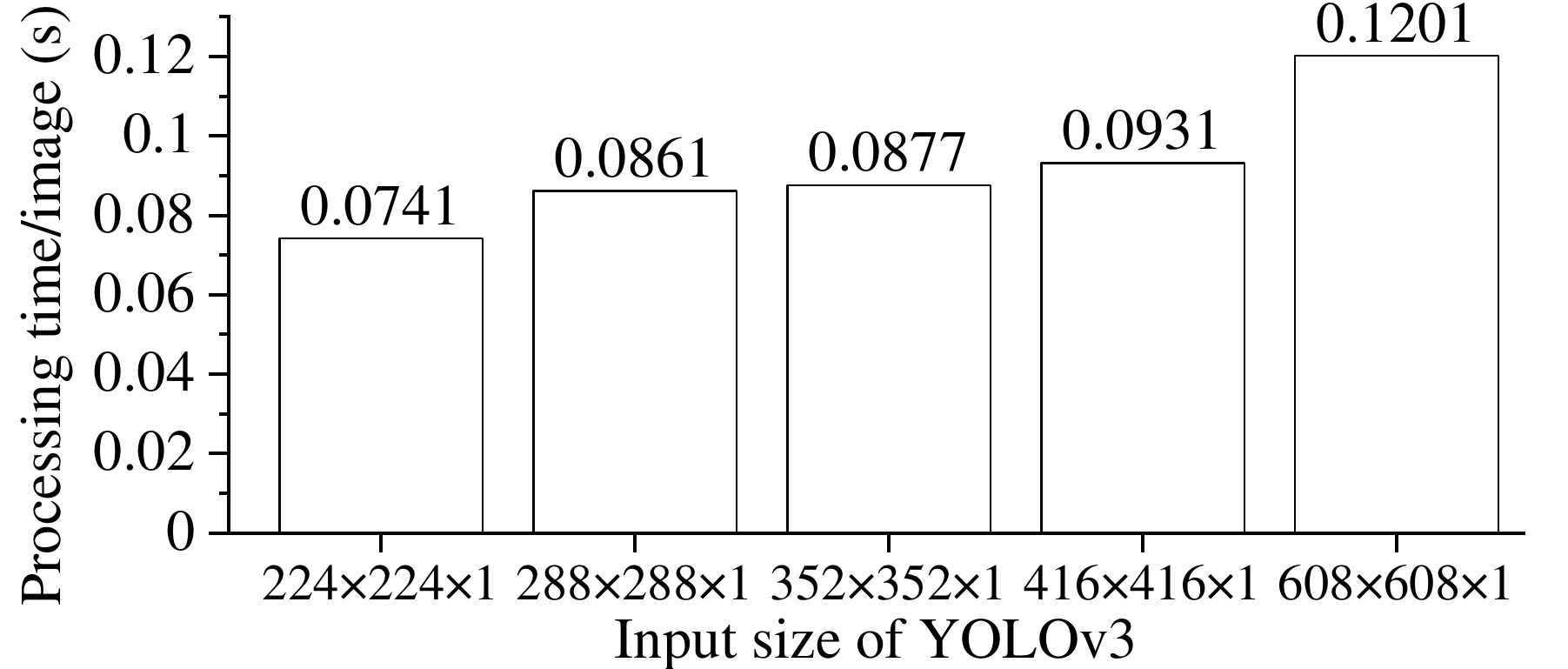}}
\caption{ The processing time per image of the YOLOv3 inference process on an NVIDIA Tesla P100 GPU.  }
\label{fig12}
\end{figure}

As is shown in Fig.~\ref{fig12}, the processing time per image of the YOLOv3 inference process varies with the input size of the DNN. Since the images are resized to the input size before fed into the DNN, the processing time of YOLOv3 inference has no relationship with the size of the image transformed from bitstream.
It is shown that the processing time of YOLOv3 inference increases as the input size of YOLOv3 increases. The model with a large input size has no accuracy improvement while sacrificing the computation cost. Besides, the processing time of inference process for SSD300 is 0.0840s per image. The processing time for inference is measured on an NVIDIA Tesla P100 GPU. The confidence threshold is set as 0.5 and the IoU threshold is set as 0.45 for inference.

\subsection{Summary}\label{4_7_sec_recap}

Based on the above experimental results and analysis, the effectiveness of the proposed bitstream function block detection methodology is proved. The following points summarize the insights from the experimental results:

1) Similar bitstream format rules can be found in several FPGA devices by bitstream format analysis, which is the basis of the transformation from bitstream to image. The image can reflect the adjacency between the CLBs in the programmable logic and has a suitable size without losing useful information. The single-channel binary image representation has the best performance.

2) The deep learning-based object detection algorithm has the capability to detect the function blocks with the same constructions in the training set, no matter whether the system designs appear in the training set or not.

3) The model trained on the bitstreams generated by one version of Xilinx Vivado can also detect function blocks from the bitstreams generated by other versions of Xilinx Vivado.

4) In the methodology, other deep learning-based object detection algorithms with high performance can also be chosen to detect function blocks from the bitstream.

\section{Conclusions}\label{6_sec_conclusions}
In this paper, we have proposed an FPGA bitstream function block detection method built upon the deep learning techniques.
At first, the bitstream format of an FPGA design was analyzed and the mapping relationship between the configuration bits and the CLB elements was found.
Then, the bitstreams of multiple FPGA designs were transformed into images suitable for deep learning processing; such transformation reflects the adjacency of the programmable logic. All these images form a dataset. In our experiment, we created a large image set from various bitstreams (for a total of 18,268 images). This dataset was then used to train the object detection algorithm based on deep learning, and upon completion of training, the algorithm is readily applied to detect function blocks from FPGA bitstreams. The entire process of dataset generation, deep learning training and testing could be fully automated.
Experimental results show that mAP (IoU=0.5) for 10 kinds of function blocks is as high as 98.11\% when using YOLOv3 as the object detector. The detector was found to have the capability to detect the function blocks from the bitstreams implementing the system designs not appearing in the training set or from the bitstreams generated by other EDA tools.

\bibliographystyle{IEEEtran}
\bibliography{bib/IEEEabrv,bib/my}

%

\begin{IEEEbiography}[{\includegraphics[width=1in,height=1.25in,clip,keepaspectratio]{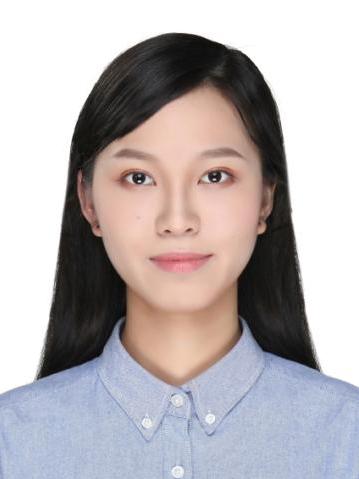}}]{Minzhen Chen}
received the B.S. degree in information and communication engineering from the college of information science and electronic engineering, Zhejiang University, Hangzhou, China, in 2018, where she is currently working towards the M.S. degree in information and communication engineering.
Her research interests include system security, hardware security, and artificial intelligence.
\end{IEEEbiography}

\begin{IEEEbiography}[{\includegraphics[width=1in,height=1.25in,clip,keepaspectratio]{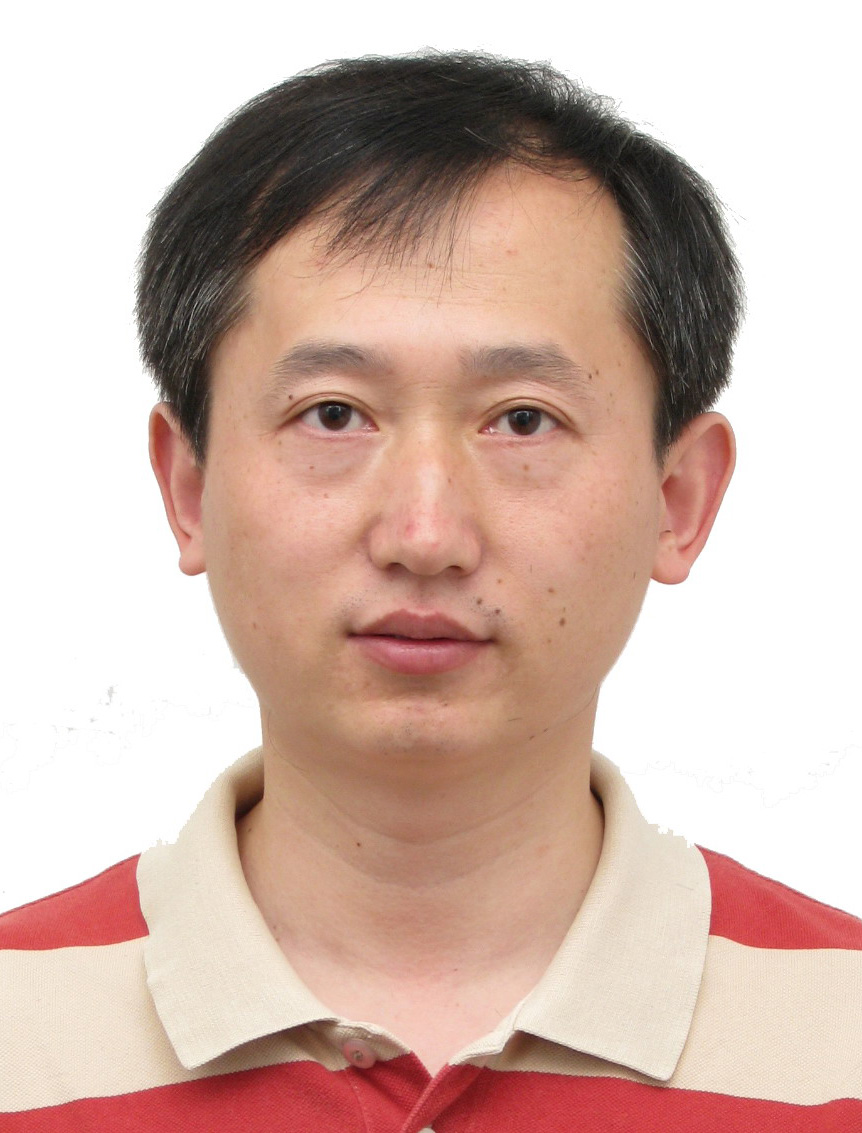}}] {Peng Liu} received the B.S. degree in optical engineering,
the M.S. degree in optical engineering, and the Ph.D. degree in communication and electronic system from Zhejiang University,
Hangzhou, China, in 1992, 1996, and 1999, respectively.
In 1999, he joined the faculty of the information science and electronic engineering department, Zhejiang University.
From 2009 to 2010, he was a Visiting Scholar with the University of Rochester, Rochester, NY, USA,
involved in high-performance computer architecture.
His current research interests include computer architectures, parallel computer architectures, VLSI design, and hardware security.
He is a member of the IEEE and the IEEE Computer Society and the IEEE Solid-State Circuits Society.

\end{IEEEbiography}

\vfill




\EOD

\end{document}